\documentclass[runningheads]{llncs}

\usepackage{eccv}

\usepackage{eccvabbrv}

\usepackage{graphicx}
\usepackage{booktabs}
\usepackage{tabularx}
\usepackage{epsfig}
\usepackage{graphicx}
\usepackage{amsmath}
\usepackage{amssymb}
\usepackage{colortbl}
\usepackage{multirow}
\usepackage{color}
\usepackage{setspace}
\usepackage[normalem]{ulem}
\usepackage{tikz}
\usepackage{comment}
\usepackage{algorithm}
\usepackage{algpseudocode}
\usepackage{float}
\usepackage{caption}
\usepackage{adjustbox}
\usepackage{pifont}%
\usepackage{wrapfig}

\definecolor{blgray}{gray}{0.97}
\definecolor{mygray}{gray}{.93}

\definecolor{top1}{RGB}{245,152,153}
\definecolor{top2}{RGB}{253,205,154}
\definecolor{top3}{RGB}{248,244,140}

\usepackage[accsupp]{axessibility}  %

\definecolor{blgray}{gray}{0.97}
\definecolor{mygray}{gray}{.93}

\usepackage{amsmath}

\usepackage{hyperref}

\usepackage[capitalize]{cleveref}
\crefname{section}{Sec.}{Secs.}
\Crefname{section}{Section}{Sections}
\Crefname{table}{Table}{Tables}
\crefname{table}{Tab.}{Tabs.}

\usepackage{orcidlink}

\begin{document}

\title{Modeling and Driving Human Body Soundfields through Acoustic Primitives}

\titlerunning{Acoustic Primitives}

\author{Chao Huang\inst{1} \and
Dejan Markovi\'{c}\inst{2} \and
Chenliang Xu\inst{1}
\and Alexander Richard\inst{2}}

\authorrunning{C. Huang et al.}

\institute{University of Rochester, Rochester, NY, USA \and
Codec Avatars Lab, Meta, Pittsburgh, PA, USA\\
\email{\{chaohuang,chenliang.xu\}@rochester.edu,\{dejanmarkovic,richardalex\}@meta.com}}

\maketitle

\begin{abstract}
While rendering and animation of photorealistic 3D human body models have matured and reached an impressive quality over the past years, modeling the spatial audio associated with such full body models has been largely ignored so far.
In this work, we present a framework that allows for high-quality spatial audio generation, capable of rendering the full 3D soundfield generated by a human body, including speech, footsteps, hand-body interactions, and others.
Given a basic audio-visual representation of the body in form of 3D body pose and audio from a head-mounted microphone, we demonstrate that we can render the full acoustic scene at any point in 3D space efficiently and accurately.
To enable near-field and realtime rendering of sound, we borrow the idea of volumetric primitives from graphical neural rendering and transfer them into the acoustic domain.
Our acoustic primitives result in an order of magnitude smaller soundfield representations and overcome deficiencies in near-field rendering compared to previous approaches. 
Our project page: \url{https://wikichao.github.io/Acoustic-Primitives/}.

\end{abstract}

\section{Introduction}

Learning, rendering, and animating 3D human body representations has been a long standing research area with applications in gaming, movies, and more recently also AR/VR.
MetaHumans~\cite{metahuman} and Codec Avatars~\cite{bagautdinov2021driving} provide highly realistic models and advances in neural rendering have pushed the visual quality to new frontiers~\cite{zielonka2023drivable,saito2023relightable,qian2023gaussianavatars}.
Animating full-body models has seen significant progress with the availability of generative models, ranging from pose-based animation~\cite{bagautdinov2021driving} to audio- and text-driven animation~\cite{ng2024audio2photoreal, tevet2023human, yi2023generating}.
Overall, \textit{visual} representations of 3D humans these days are of excellent quality and drivable from pose, audio, and text inputs.

However, on the \textit{acoustic} side of the problem, \ie, rendering spatial sound in 3D for these full-body representations, the research landscape looks dire.
It has been shown that accurate audio-visual modeling is important for an immersive 3D experience~\cite{hendrix1996sense} but still almost no research exists that would allow to render spatial audio of virtual humans.
Analogous to visual full-body models, acoustic full-body models have similar requirements:
first, it must be possible to render spatial sounds produced by a virtual human at any position in 3D space, and second, the soundfield needs to be drivable. In this work, we focus on generating and driving full-body soundfields from 3D body pose and head-mounted microphones.

\begin{figure}[t!]
    \centering
    \includegraphics[width=0.99\textwidth]{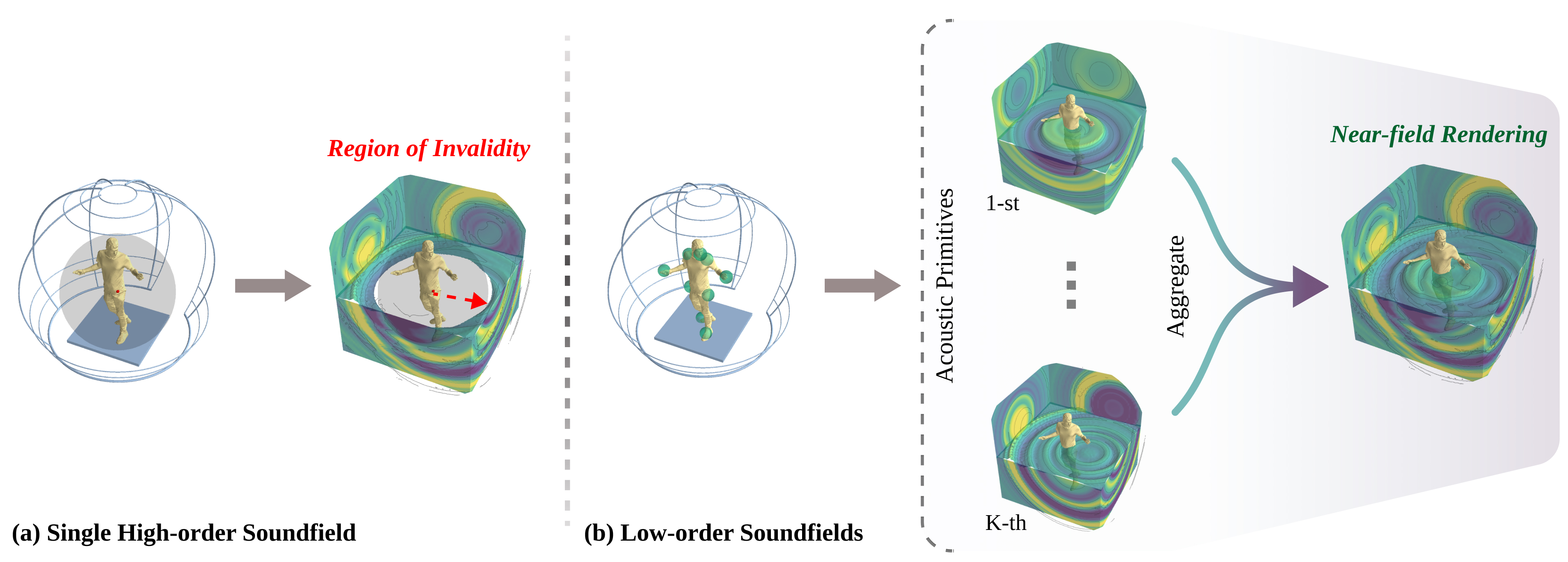}
    \caption{\textbf{Single high-order soundfield (a) vs. acoustic primitives (b)}. Existing approach~\cite{xudong2023sounding} predicts a high-order ambisonic soundfield around the human body, preventing sound from being rendered in the near-field; our proposed acoustic primitives, represented as small spheres attached to the body, successfully model a complete and accurate 3D body soundfield.}
    \label{fig:overview}
\end{figure}

This problem has recently been addressed in pioneering work by Xu \etal~\cite{xudong2023sounding}, who developed a neural soundfield rendering system for full body avatars, driven by body pose and headset microphone input.
However, \cite{xudong2023sounding} has several major limitations:
The approach relies on a single high order ambisonic (spherical harmonics) representation that models the sound emitted from the surface of a sphere around the human body, with a diameter of about 2m.
Sound can only be modeled outside of this sphere, such that near-field modeling of signals closer to the body is not possible, see \cref{fig:overview}a.
Moreover, accurate sound reproduction in~\cite{xudong2023sounding} relies on extremely high-order ambisonic coefficients which are expensive to compute and instable to estimate.
To get around this instability, \cite{xudong2023sounding} does not predict the ambisonic coefficients directly, but instead predicts the raw audio signal on 345 positions surrounding the body, and then uses traditional signal processing to compute a 17-th order ambisonic representation from these 345 raw waveforms.
This mechanism is computationally inefficient and prevents realtime sound rendering.

In this work, we propose a novel sound rendering method based on acoustic primitives which solves the problems of~\cite{xudong2023sounding}:

\noindent
\textit{Near-field Rendering.}
We take inspiration from recent methods in visual neural rendering that rely on \textit{volumetric primitives} like cuboids~\cite{lombardi2021mixture} or Gaussians~\cite{kerbl3Dgaussians}.
Instead of modeling the body soundfield by an ambisonic representation on a single sphere around the full human body as in~\cite{xudong2023sounding}, we attach multiple acoustic primitives (small spheres each representing low-order ambisonics) to the 3D human skeleton and model the sound radiating from each of these acoustic primitives separately, see ~\cref{fig:overview}b.
The full soundfield produced by all primitives together is given by the sum of the individual rendered sound from each primitive.
This way, sound can be modeled arbitrarily close to the body.

\noindent
\textit{Efficient Soundfield Representation.}
Instead of a single 17-th order ambisonic representation, we model body sounds by multiple low-order (typically second order) ambisonic primitives.
This reduces the number of parameters characterizing the acoustic scene by an order of magnitude and allows for a more compact and efficient soundfield representation.

\noindent
\textit{Efficient Rendering.}
Instead of predicting 345 raw audio signals and relying on traditional, costly high-order ambisonic encoders and decoders, we predict the low-order ambisonic coefficients of each primitive directly.
Efficient sound rendering can then be achieved using spherical wave functions as described in \cref{subsec:spherical_wave_functions}.

\noindent
\textit{Drivability.}
Same as~\cite{xudong2023sounding}, our method can be driven from body pose and a head mounted microphone, \ie 3D soundfields can be generated for novel acoustic input and body motion.
Note that this is in stark contrast to its visual counterparts~\cite{kerbl3Dgaussians,mildenhall2020nerf} which are designed to synthesize novel views of fixed scenes, but are typically not drivable from user input.

In summary, we propose an efficient and drivable 3D sound rendering system with
\begin{enumerate}
    \item \textbf{audio-visual driving:} given body pose and an audio signal from head-mounted microphones, we can accurately render the soundfield produced by the body (speech, snapping, clapping, footsteps, etc) in 3D;
    \item \textbf{real-time rendering:} the introduction of acoustic primitives allows for efficient real-time rendering of 3D sound scenes;
    \item \textbf{high quality:} although relying only on low-order ambisonic representations, we achieve comparable quality to~\cite{xudong2023sounding} but avoid the high computational cost.
\end{enumerate}

\section{Related Works}

\noindent
\textbf{Spatial Audio Modeling.}
Existing works on spatial audio rendering are either based on traditional signal processing and linear filters~\cite{savioja1999creating,chen2020soundspaces,chen2022soundspaces2} or on more recent neural binaural renderers~\cite{richard2021neural,richard2022deepimpulse,gebru2021implicit}.
While these approaches can typically produce spatial audio in an efficient way, they come at strong restrictions, particularly, they need to know the exact location of each sound source to render as well as the clean sound signal for each sound location.
Such information is available in fully synthetic, artist-created scenes but is usually unknown in real environments and real acoustic scenes.
Our approach, in contrast, does not rely on such knowledge and implicitly learns to separate an aggregated audio signal into its distinct sources (the acoustic primitives) through inverse acoustic rendering.
More recently, data-driven methods aim to produce binaural audio from audio-visual input information.
Chen \etal~\cite{chen2023novelview} propose a system to render a pre-recorded 3D acoustic scene from novel viewpoints, however, this method can not handle new acoustic scenes. Liang \etal~\cite{liang23avnerf} propose to reconstruct the 3D audio-visual scene from videos, but the scene is static.
Gao \etal~\cite{gao20192} analyze a 2D image of a visual scene to generate binaural audio for a given monaural sound source. Note that this approach can only deal with single sound sources and can't handle complex 3D sound scenes.
In~\cite{morgado2018self}, the authors propose a method to generate spatial audio from mono acoustic input and 360 degree camera inputs. They demonstrate that their approach can correctly localize sound sources in the scene and generate correct spatial audio at a coarse resolution.

\noindent
\textbf{Primitives in Volumetric Rendering.}
Neural rendering has been revolutionized by volumetric rendering methods like neural volumes~\cite{lombardi2019neuralvolumes} and neural radiance fields~\cite{mildenhall2020nerf}.
Follow-up works build on volumetric primitives such as cuboids~\cite{lombardi2021mixture} or Gaussians~\cite{kerbl3Dgaussians} and render via ray-marching or splatting.
These technologies have unlocked real-time rendering for animatable avatars~\cite{chen2023primdiffusion,lombardi2021mixture,remelli2022drivable,zielonka2023drivable,saito2023relightable}.
We borrow the idea of volumetric primitives and transfer them from the visual domain into the acoustic domain to build an efficient and low-parameter characterization of soundfields.

\noindent\textbf{Audio-Visual Learning.}
Audio-visual learning has been widely applied to find connections between acoustic and 2D visual signals, \eg in audio-visual localization~\cite{qian2020multiple,hu2020discriminative,tian2021cyclic,jiang2022egocentric,mo2022localizing,huang2023egocentric}, for source-separation~\cite{ephrat2018looking,zhao2018sound,gao2018learning,gao2021visualvoice,huang2023davis}, or to learn associations from 360-degree videos~\cite{morgado2018self,li2018scene,huang2019audible}.
Application of audio-visual learning to 3D settings is mostly limited to 3D visual scenarios, such as audio-visual driving of avatars~\cite{yi2023generating,ng2024audio2photoreal,richard2021meshtalk,xing2023codetalker,richard2021audiogaze} or audio-driven gesture synthesis~\cite{ginosar2019learning,lee2019dancing,ahuja2020style,yoon2020speech}.
While these works use audio-visual input to learn information about a scene, they operate on visual outputs and do not model acoustic scenes.

Most closely related to our work is~\cite{xudong2023sounding}, who address the same task we address in this work.
However, as outlined above, \cite{xudong2023sounding} has some significant drawbacks such as the inability to render near-field audio, or the lack of real-time rendering capabilities.

\section{Pose-Guided Soundfield Generation using Acoustic Primitives}

\subsection{Problem Definition}
Let $ \mathbf{a}_{1:T_a} (a_1,\dots,a_{T_a}) $ be the input audio signal from one or multiple head-mounted microphones, and $ \mathbf{p}_{1:T_p} = (p_1,\dots,p_{T_p}) $ be the corresponding sequence of 3D body pose, where each $ p_t \in \mathbb{R}^{J \times 3} $ is a vector containing 3D body joint coordinates.
We aim to predict a sound signal $ \mathbf{s} $ at an arbitrary 3D position $ (r, \theta, \varphi) $. Note that we use spherical coordinates.
In other words, we learn to aim a mapping from 3D body pose, headset audio signal, and a 3D position in space to the audio signal at that 3D spatial position,
\begin{align}
    \mathbf{a}_{1:T_a}, \mathbf{p}_{1:T_p}, (r, \theta, \varphi) \mapsto \mathbf{s}_{1:T_a}.
    \label{eq:mapping}
\end{align}

The core challenge is how to get training data to learn such a model.
It is impossible to place microphones at all positions in 3D space to get dense sampling of the space.
Instead, we follow the strategy of~\cite{xudong2023sounding} (and actually use the same public dataset for our work) and sample soundfield signals $ \mathbf{s}_{1:T_a} $ only on a sphere around the human body.
This poses the challenge of rendering the soundfield at positions that are not on the surface of the sphere on which data has been captured.
Note the analogy to graphics neural rendering: approaches like NeRF~\cite{mildenhall2020nerf} also don't have spatially dense samples of 2D images taken from a scene, yet through inductive biases such as the rendering equation, they succeed in synthesizing the 3D scene from any novel viewpoint.
We apply the same strategy for audio, and learn the mapping in Eq.~\eqref{eq:mapping} by differentiation through the wave propagation function, which we explain in the next section.

\subsection{Sound Radiation using Spherical Wave Functions}
\label{subsec:spherical_wave_functions}

The general solution of the homogeneous, time-dependent wave equation in the spherical coordinate system is given by 
\cite{williams1999fourier, zotter2019ambisonics}:
\begin{equation}
\mathbf{w}(t, f, r,\theta,\varphi) = \sum_{n=0}^{\infty} \sum_{m=-n}^{n} \left[  b_{nm}(t, f) \cdot j_n (kr) + c_{nm}(t, f) \cdot h_n (kr) \right] \cdot Y_{nm} (\theta, \varphi), \\
\label{eq:wave_equation}
\end{equation}
where $(r, \theta,\varphi)$ are arbitrary coordinates inside a {\bf source-free region}, $t$ and $f$ denote the time and frequency (we will omit them in the following sections for clarity), and $k = 2\pi f / v_{sound}$ is the corresponding wavenumber; $Y_{nm} (\theta, \varphi)$ represents the spherical harmonic of order $n$ and degree $m$, and $j_n (kr)$ and $h_n (kr)$ are, respectively, $n$th-order spherical Bessel and Hankel functions. Coefficients $b_{nm}(t, f)$ and $c_{nm}(t, f)$ describe, respectively, incoming  and outgoing waves. In particular, considering the scenario depicted in \cref{fig:overview}a, only the radiating field component is present, \ie $b_{nm} = 0$, which in literature is known as exterior domain problem \cite{williams1999fourier, samarasinghe2012spatial}. 

Given recorded or predicted microphone signals on the surface of the sphere surrounding the body, SoundingBodies~\cite{xudong2023sounding} approaches the sound field modeling task as a traditional exterior domain problem and estimates the sound field coefficients $c_{nm}(t, f)$. While the general solution requires an infinite number of harmonic orders, the practical estimates are limited by the available number of microphones $M$ as $N =\sqrt{M}-1$,
\begin{equation}
\hat{\mathbf{w}}(r,\theta,\varphi) = \sum_{n=0}^{N} \sum_{m=-n}^{n} \hat{c}_{nm} \cdot h_n (kr)  \cdot Y_{nm} (\theta, \varphi). \\
\label{eq:exterior_problem}
\end{equation}
This reliance on a generic solution of the exterior domain problem limits the practicability of \cite{xudong2023sounding}. While the network uses pose information to predict microphone signals, the successive DSP processing of these signals required for spatial sound rendering (\eg binauralization) does not leverage pose information at all. 
This brings two main issues: 
\begin{enumerate}
  \item To model sound sources located further away from the center of the representation, see $R_0$ in \cref{fig:overview}a, higher harmonic orders are needed, which in turn requires the network to predict a high number of microphones before any spatial rendering can be performed; predicting a smaller number of signals would limit the harmonic order and, consequently, the rendered scene would collapse towards the center.
  \item The wave equation solution is valid only outside the boundary surface encompassing all sources of sound, as shown in \cref{fig:overview}a. Inside this region, the high harmonic orders produce chaotic results, limiting the minimum distance at which the scene can be rendered and experienced.
\end{enumerate}
To address the above issues we take a different approach. Instead of using pose-conditioned network to predict microphone signals, we use the network to predict sound field coefficients directly. Furthermore, instead of trying to estimate a generic high-order sound field representation, we leverage the knowledge of possible positions of sound, given by the body pose, and model the sound radiation as a superposition of several small-order elementary sound fields originating from different positions of the body as depicted in 
\cref{fig:overview}b. Similarly to Eq.~\eqref{eq:exterior_problem}, the sound pressure produced by a single elementary field of order $N$ is given by 
\begin{equation}
\begin{split}
\mathbf{w}(r,\theta,\varphi) & = \sum_{n=0}^{N} \sum_{m=-n}^{n} \left({c}_{nm} \cdot h_n (kr_{ref})\right) \cdot \frac{h_n(kr)}{h_n(kr_{ref}) }  \cdot Y_{nm} (\theta, \varphi) \\
& = \sum_{n=0}^{N} \sum_{m=-n}^{n} \Tilde{c}_{nm} \cdot \frac{h_n (kr)}{h_n(kr_{ref}) }  \cdot Y_{nm} (\theta, \varphi),
\end{split}
\label{eq:primitives}
\end{equation}
where we use $h_n(kr_{ref})$ with $r_{ref}=0.5\mathrm{m}$ for numerical stability of the learning process, and refer to harmonic coefficients $\mathcal{S} = [\Tilde{c}_{00},...,\Tilde{c}_{NN}]$ as an \textbf{acoustic primitive}. 

Given Eq.~\eqref{eq:primitives}, we can translate the task of modeling 3D spatial sound for the visual body to learning a set of small acoustic primitives $\{\mathcal{S}_i\}_{i=1}^{K}$, which we choose $N$ up to the second order for harmonic coefficients and set the number of acoustic primitives as $K$. In practice, capturing ground truth sound field coefficient is infeasible, while the microphone signals received on the surface of the dome are available with prior efforts by \cite{xudong2023sounding}. Since the produced sound pressure $\mathbf{w}(r,\theta,\varphi)$ indeed represents the audio signal produced at spherical position $(r,\theta,\varphi)$, we can therefore decompose the entire learning process into two sub-steps:

\begin{itemize}
    \item \noindent\textbf{Learning Acoustic Primitives.} 
The main objective of this step is to design a neural network $\mathcal{F}$ that consumes audio and pose data as input, and output the sound field representation
\begin{equation}
    \{\mathcal{S}_i\}_{i=1}^{K} = \mathcal{F}(\mathbf{a}_{1:T_a}, \mathbf{p}_{1:T_p}).
\end{equation}

\item  \noindent\textbf{Rendering Audio with Learned Acoustic Primitives.} With the learned acoustic primitives $\{\mathcal{S}_i\}_{i=1}^{K}$, we leverage Eq.~\eqref{eq:primitives} as a differentiable rendering function, denoted as $\mathcal{R}$, to generate the audio waveform received at the target position
\begin{equation}
    \hat{\mathbf{s}}_{1:T_a}(r, \theta,\varphi) = \mathcal{R}(\{\mathcal{S}_i\}_{i=1}^{K}, r, \theta,\varphi).
    \label{eq:waverenderer}
\end{equation}
\end{itemize}

The hyperparameter in $\mathcal{R}$ is fixed once the harmonic order $N$ and the number of primitives $K$ are initialized, and all the operations in $\mathcal{R}$ are differentiable, making it feasible to run end-to-end training. With the training data tuple $\left(\mathbf{a}_{1:T_a}, \mathbf{p}_{1:T_p}, \mathbf{s}_{1:T_a}(r, \theta,\varphi)\right)$ that includes recorded microphone signal at position $(r, \theta,\varphi)$, we can learn a pose-guided acoustic primitive synthesis system by simply optimizing the loss between $\hat{\mathbf{s}}_{1:T_a}$ and $\mathbf{s}_{1:T_a}$.

\begin{figure}[!t]
    \centering
    \includegraphics[width=0.99\textwidth]{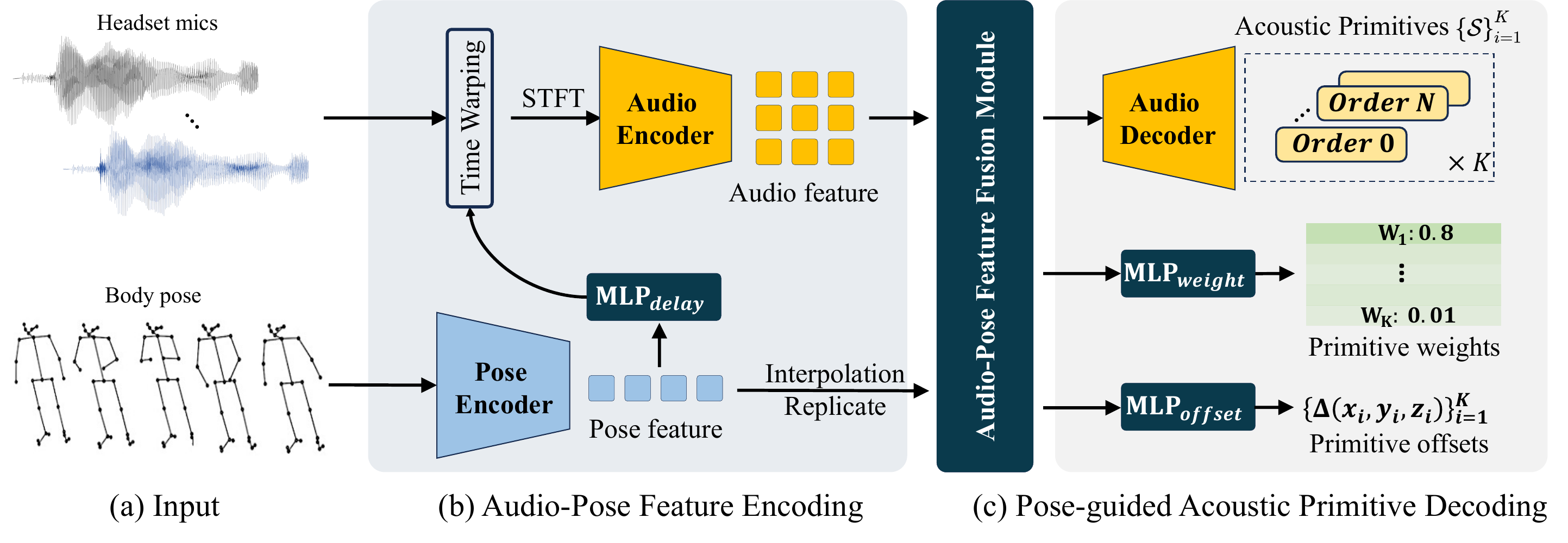}
    \caption{Our pose-guided acoustic primitive learning framework takes headset microphone signals and body pose information as inputs. It outputs the acoustic primitive representations, weights, and offsets in one pass. The framework consists of two main stages. In the first stage, we employ separate encoders to process the audio and pose signals into feature spaces. An Audio-Pose Feature Fusion Module is then utilized to combine these features. In the second stage, the fused features are fed into an audio decoder network to generate the acoustic primitive coefficients. Additionally, two separate MLP heads are used to predict the weights and offsets for each acoustic primitive.}
    \label{fig:framework}
\end{figure}

\subsection{Multimodal Feature Encoding}
\label{subsec:fusion}

\noindent\textbf{Pose Encoder.}
Human body movements offer crucial clues for how sound is distributed in space.
To capture these rich spatial cues, we employ a pose encoder that processes the input pose sequence $ \mathbf{p}_{1:T_p} $.
The pose input is first encoded into a latent feature representation. To capture temporal relationships, we apply two layers of temporal convolutions with a kernel size of 5. Finally, we concatenate the encoded features for all joints and use an MLP to create a compact representation, denoted as $f_p \in \mathbb{R}^{C_p \times T^{'}_p}$. Here $C_p$ is the number of feature channels and $T^{'}_p$ represents the temporal dimension after convolution. More details are provided in the supplementary material.

\noindent\textbf{Audio Encoder.}
While sound can originate from various points on the body (\eg, hands, feet), it's captured by the headset microphone located at a central position near the head. This difference in location creates a slight time delay between the moment the sound is produced and when it's actually recorded. Previous research has shown that compensating for this time delay can be beneficial~\cite{richard2021neural,xudong2023sounding}. In our approach, we leverage the pose features as guidance and use an MLP (as shown in \cref{fig:framework}) to estimate the delay for each acoustic primitive attached to a body joint, and time-warp the audio signal accordingly.

Via STFT, the warped signals are then transformed into complex spectrograms $X^c_a \in \mathbb{R}^{C_h \times F \times T}$ where $F$ and $T$ represent the number of frequency and time bins respectively, and $C_h$ is the number of audio channels.
The resulting audio features are then encoded with a network consisting of convolutional- and LSTM layers to capture both local context and long-range dependencies within the audio data. The encoder architecture utilizes four layers, where each layer contains two ResNet blocks, a temporal LSTM block, and a downsampling block with a factor of 2. Eventually, we can obtain the latent audio features $f_a = \mathrm{E}_a(X_a) \in \mathbb{R}^{C_a \times \frac{F}{16} \times \frac{T}{16}}$.

\noindent\textbf{Audio-Pose Feature Fusion Module.}
While headset audio reveals the content of sounds (\eg, finger snapping), it lacks precise spatial information about the source. Conversely, body pose offers strong spatial cues about joint locations, but cannot identify the sound type (\eg, speech) solely from pose data.  Therefore, effectively combining audio and pose features is crucial for learning acoustic primitives and determining their contribution to the final sound generation.
We first interpolate the pose features $f_p$ to match the temporal size of the audio features $f_a$.
We then employ a lightweight fusion module with two ResNet blocks and one attention block to combine the concatenated audio and pose features, resulting in a new representation denoted as $f_{ap} \in \mathbb{R}^{C_a \times \frac{F}{16}  \times \frac{T}{16}}$.

\begin{figure}[!t]
    \centering
    \includegraphics[width=0.99\textwidth]{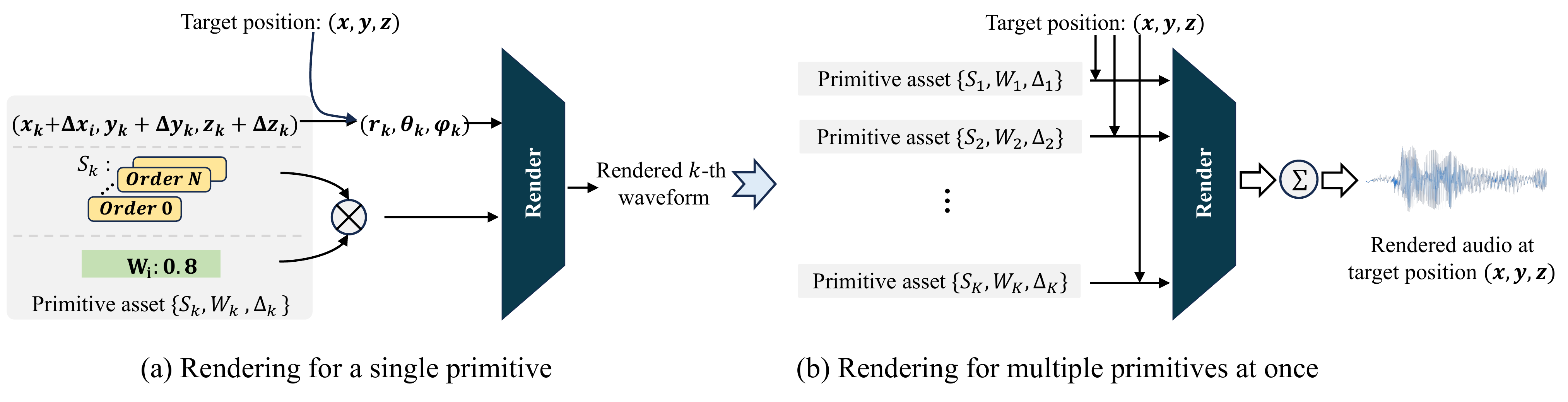}
    \caption{Illustration on the rendering process with the estimated acoustic primitives. (a) demonstrates how to render a waveform signal given the learned harmonic coefficients $\mathcal{S}_k$, primitive coordinate offset $\Delta_k$, and the weight $W_k$. Next, we show that for all the primitives, we render audio generated by the primitive at the targeted location and aggregate them to yield the final rendered audio at target position $(x,y,z)$. }
    \label{fig:rendering}
\end{figure}

\subsection{Acoustic Primitive Decoding}
\label{subsec:decoding}
As described in \cref{subsec:spherical_wave_functions}, an acoustic primitive determines the audio heard at arbitrary coordinates inside its sound field. It considers factors like the harmonic coefficients, the center coordinate of the primitive, and the target location where the sound is perceived. In this work, we focus on generating sound based on the target location. This translates to learning two key components: the primitive's coordinates and the harmonic coefficients.  However, this is non-trivial as the primitive's location changes dynamically as the body moves.  Additionally, the harmonic coefficients must capture not only the sound content but also the spatial cues such as sound directivity. In the next section, we will explain how our approach addresses these challenges.

\noindent\textbf{Sound Field Decoder.}
Leveraging the fused features $f_{ap}$, our decoder $\mathrm{D}_a$ simultaneously generates the sound field representations for all the acoustic primitives. 
Similar to the input spectrograms $X_a$, the harmonic coefficients have the same spatial dimensions but differ in the number of channels, which encode the richness of the sound's spatial information. A higher number of channels allows for more precise control over the perceived location of the sound. For simplicity, we design the decoder to resemble the audio encoder and add skip connections between the encoder and decoder, with the main difference being the number of output channels. The decoder outputs ${(N+1)}^2 \times K$ channels. Here, $N$ represents the order of the harmonic coefficients, which controls the level of detail captured in the spatial representation.  
Finally, we separate the decoder's output into $K$ distinct harmonic coefficients $\{\mathcal{S}_i\}_{i=1}^{K}$, one for each acoustic primitive.

\noindent\textbf{Primitive Offsets.}
We initialize acoustic primitives to be at body joint locations, \eg at the wrists, face and ankles.
While the initial 3D coordinates of these body joints provide a reasonable starting point, the actual locations of the sound sources might differ slightly from the body joint positions. For example, the chosen keypoints represent wrists, but the sound of finger snapping originates from the fingers themselves. This discrepancy between the body joint and sound production locations can affect the learning process and the accuracy of the rendered spatial audio. To address this limitation, we learn offsets for the initial coordinates to better represent the actual positions of acoustic primitives. In practice, we employ a three-layer MLP network that operates on the fused features $f_{ap}$. First, we apply mean pooling along the frequency axis of $f_{ap}$ but keep its time dimension, obtaining $\mathop{\bar{f_{ap}}}$. Then, we generate the offsets by
\begin{equation}
    \Delta(x,y,z) =  \sigma \cdot \tanh\left(\mathrm{MLP}_{\text{offset}}\left(\mathop{\bar{f_{ap}}}\right)\right).
\label{eq:offset}
\end{equation}
To constrain the predicted offsets within a reasonable range, we use a $\tanh$ activation function and apply a scaling factor of $\sigma = 0.2$ to restrict the offsets to a maximum range of 20 centimeters around the initial locations.

\noindent\textbf{Primitive Weights.}
At different points in time, primitives have different importance.
For instance, when a finger is snapped, the hand primitive emits high energy sound while other primitives emit at most low energy.
The relationship is a function of the input audio and body pose.
We therefore explicitly model the weight of each primitive as a function of the combined audio and pose encodings $\mathop{\bar{f_{ap}}}$,
\begin{equation}
    W = \mathrm{softmax}\left(\mathrm{MLP}_{weight}\left(\mathop{\bar{f_{ap}}}\right)\right).
\label{eq:weight}
\end{equation}
$W$ is the predicted weight for each primitive at each time instance and indicates the relative influence in the final rendered sound.

\subsection{Differentiable Acoustic Primitive Renderer}
\label{subsec:rendering}
Given the initial primitive locations (\ie the joint locations to which the primitives are attached), the learned offsets, harmonic coefficients, and weights, we can now render the sound field for each primitive (as shown in 
\cref{fig:rendering}).
We first compute the primitive's predicted location by adding the learned offsets to the corresponding body joint location.
Now, given a listener position in 3D space at which we want to render the sound, we transform each primitive's predicted location into spherical coordinates $(r_k, \theta_k, \varphi_k)$ representing the relative position of the listener with respect to each of the $K$ primitives.
We now use the differentiable audio renderer from Eq.~\eqref{eq:waverenderer} to render the audio signal $ \mathbf{\hat{s}}_{1:T_a}^k $ produced by the $k$-th primitive at the listener's position,
\begin{align}
    \mathbf{\hat{s}}_{1:T_a}^k = \mathcal{R}(\mathcal{S}_k \cdot W_k, r_k, \theta_k, \varphi_k),\
\end{align}
and obtain the full sound field by summation over all acoustic primitives,
\begin{align}
    \mathbf{\hat{s}}_{1:T} = \sum_k^{K} \mathbf{\hat{s}}_{1:T}^k.
\end{align}

\subsection{Loss Function}
\label{subsec:loss}
Since our renderer $\mathcal{R}$ is differentiable, it allows us to efficiently train the model using loss functions on the final predicted waveforms. In this work,
we employ a multiscale STFT loss~\cite{yamamoto2021parallel} between the predicted audio $\mathbf{\hat{s}}_{1:T_a}$ and the ground truth audio $\mathbf{s}_{1:T_a}$ on their amplitude spectrograms, denoted as $\mathcal{L}_{amp}(\mathbf{\hat{s}}_{1:T_a},\mathbf{s}_{1:T_a})$ and on the real and imaginary parts of spectrograms, denoted as $\mathcal{L}_{ri}(\mathbf{\hat{s}}_{1:T_a},\mathbf{s}_{1:T_a})$. The window sizes are set as 2048, 1024, 512, 256. As proposed in \cite{xudong2023sounding}, a shift-$\ell1$ loss helps reduce the spatial alignment error. We therefore add this loss term as $\mathcal{L}_{s\ell1}(\mathbf{\hat{s}}_{1:T_a},\mathbf{s}_{1:T_a})$. Additionally, determining a primitive's contribution to the final sound (corresponding to the primitive weights $W$) can be challenging without additional guidance. To overcome this, we leverage clip-level labels (denoted by $\mathrm{y} \in \mathbb{R}^K$) that specify which body joint contributes to the received audio. We apply average pooling along the frequency dimension and find the maximum value of $W$ for each primitive across all time steps, resulting in $\Tilde{W} \in \mathbb{R}^K$. This essentially summarizes whether an acoustic primitive has contributed to the sound in the audio clip. Finally,  a simple cross-entropy loss function $\mathcal{L}_{cts}(\Tilde{W}, \mathrm{y})$ is employed to aid in the learning process. Our final loss becomes
\begin{equation}
    \mathcal{L}_{total} = \lambda_{amp}\mathcal{L}_{amp}+\lambda_{ri}\mathcal{L}_{ri}+\lambda_{s\ell1}\mathcal{L}_{s\ell1}+\lambda_{cts}\mathcal{L}_{cts}.
\label{eq:loss}
\end{equation}
Please refer to the supplementary materials for ablation on the loss terms.

\section{Experiments}
\label{sec:experiments}

\subsection{Experimental Setting}
\label{subsec:setting}

\noindent\textbf{Dataset.} To evaluate our approach, we leverage the publicly available dataset introduced in ~\cite{xudong2023sounding} \footnote{\url{https://github.com/facebookresearch/SoundingBodies}}\footnote{Note: data used in the paper and data released publicly differ by 1.5 subjects (8 subjects used in \cite{xudong2023sounding} vs 6.5 publicly released). We updated the performance metrics of the baseline \cite{xudong2023sounding} to account for this difference.}. The dataset captures synchronized audio and visual data in an anechoic chamber, offering multimodal data specifically designed for speech and body sound field modeling research. It utilizes 5 Kinect sensors for body tracking and a large microphone array (345 microphones) arranged in a spherical fashion around the recording area. The data encompasses various participants performing a diverse range of body sounds and speech in different settings (\eg, standing or sitting). The recordings are segmented into non-overlapping one-second clips.
We adopt the same train/validation/test splits established by~\cite{xudong2023sounding}, resulting in 10,076/1,469/1,431 clips, respectively.

\noindent\textbf{Implementation Details.} In our experimental setup, we employ a sampling rate of 48 kHz for audio signals and a frame rate of 30 fps for body pose data. The audio waveforms are converted into complex spectrograms using a Hann window of size 512 and a hop length of 128 and FFT length of 1022. Within the encoders, both the pose features $f_p$ and audio features $f_a$ are configured to have the same channel size $C_a = C_p = 256$. We set the order of harmonic coefficients to $N=2$. During training, the batch size is set as 1 per GPU and we randomly select 20 microphones from the available pool of 345 target microphones for each forward pass. The AdamW optimizer with a learning rate of 0.0002 is used, and the network is trained for 100 epochs. To balance different loss terms, we set the weights $\lambda_{amp}=7$, $\lambda_{ri}=3$, $\lambda_{s\ell1}=0.5$, and $\lambda_{cts}=1$. The experiments are conducted on 4 NVIDIA Tesla A100 GPUs, with model training for 100 epochs taking approximately 55 hours to complete.

\noindent\textbf{Evaluation Metrics.} We evaluate the performance of our model using three main metrics: the signal-to-distortion ratio (SDR), the $\ell_2$ error on the amplitude spectrogram, and the angular error of the phase spectrogram. The SDR measures the overall quality of the reconstructed sound, with higher values indicating better quality. The amplitude error shows how well the reconstructed sound matches the original in terms of the distribution of sound energy, while the angular error evaluates the timing accuracy of the reconstructed sound waves relative to the original. We report amplitude errors multiplied by a factor of 1000 to remove leading zeros.

\begin{table}[!t]
    \caption{Our method achieves \textbf{comparable results} to the baseline SoundingBodies~\cite{xudong2023sounding} in SDR, amplitude error, and phase error, while significantly outperforming the baseline in inference speed (\textbf{15x faster}). 
    }
    \centering
    \footnotesize
    \begin{tabularx}{1\textwidth}{lrXrrrXrrr}
        \toprule
                        & & & \multicolumn{3}{c}{non-speech}                                        & & \multicolumn{3}{c}{speech} \\
                            \cmidrule(lr){4-6}                                                        \cmidrule(lr){8-10}
        Methods    & \ \ \ Speed & & SDR $\uparrow$    & amplitude $\downarrow$    & phase $\downarrow$    & & SDR $\uparrow$    & amplitude $\downarrow$    & phase $\downarrow$ \\
        \midrule

        ~\cite{xudong2023sounding}             & {3.56s} & & {$3.052$}  & {$0.832 $}        & {$0.314 $}    & & {$9.635$}  & {$0.701 $}        & {$0.464 $}\\
        Ours & \textbf{{0.24s}}& & {$3.597$} & {$0.883$} & {$0.323$} & & {$8.448$} & {$0.943$}  & {$0.417$} \\
        
        \bottomrule
    \end{tabularx}
    \label{tab:method_compare}
\end{table}

\subsection{Comparison with Baseline}
\label{subsec:comparison}
We compare our method using $12$ acoustic primitives of $2$nd order with the SoundingBodies~\cite{xudong2023sounding} baseline. Results are shown in ~\cref{tab:method_compare}. We can observe that the sound field modeling performance of the proposed method is comparable to ~\cite{xudong2023sounding} while having a much faster inference speed. In particular, proposed method even performs better than the baseline on SDR metric for non-speech sounds and phase metric for speech. Regarding the inference speed, we show average time needed to compute 1 second of audio at 48 kHz. Note that for \cite{xudong2023sounding} we only report the time needed for the network to predict the microphone signals. In a practical scenario \cite{xudong2023sounding} needs also DSP processing of these microphone signals to obtain the high-order sound field representation, which would further increase the overall processing overhead.

\begin{figure}[!t]
    \centering
    \includegraphics[width=0.99\textwidth]{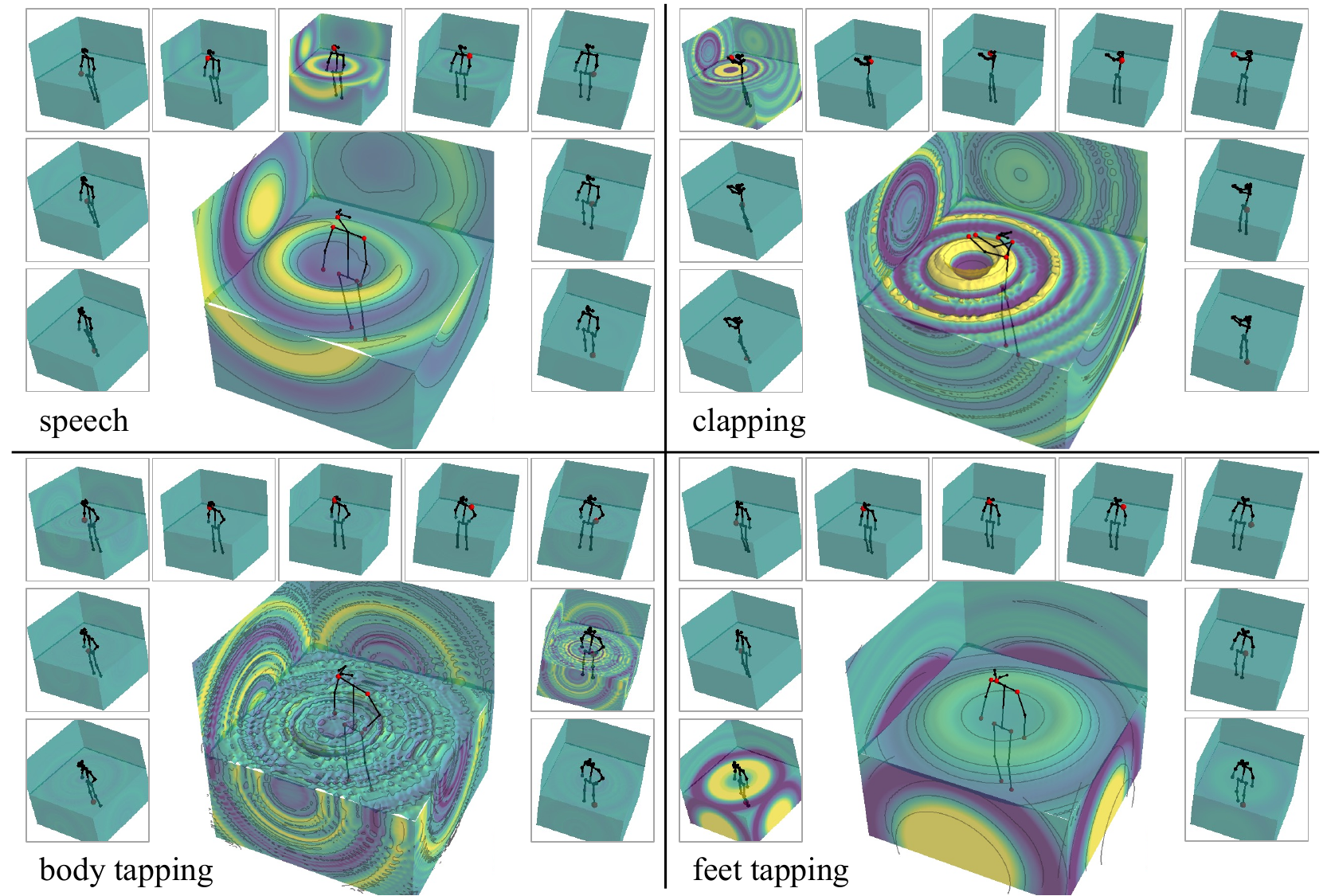}
    \caption{Sound field visualizations for 4 different kinds of sound. Main sound field is in the center and individual primitive contributions are shown around. We can observe that the method assigns acoustic energy to correct acoustic primitives, e.g. speech comes mostly from the head with only a very small contribution from the shoulder primitives. We can also observe the speech directivity pattern matching the head orientation.  For each visualization, the left/right 4 primitives are labeled as follows: \textit{foot}, \textit{hip}, \textit{hand}, and \textit{shoulder} (from bottom to top), and the middle one is the head.}
    \label{fig:examples_soundfield}
\end{figure}

\begin{table}[!t]
    \caption{Ablation study: the number and harmonic order of acoustic primitives.}
    \centering
    \footnotesize
    \begin{tabularx}{1\textwidth}{llllXrrrXrrr}
        \toprule
                    \multicolumn{2}{l}{\multirow{2}{*}{\# $K$}} & \multicolumn{2}{r}{\multirow{2}{*}{\# $N$}} & &  \multicolumn{3}{c}{non-speech}                                        & & \multicolumn{3}{c}{speech} \\
                           \cmidrule(lr){6-8}                                                        \cmidrule(lr){10-12}
        \multicolumn{2}{l}{} & \multicolumn{2}{l}{} & & SDR $\uparrow$   & amp. $\downarrow$    & phase $\downarrow$    & & SDR $\uparrow$    & amp. $\downarrow$    & phase $\downarrow$ \\
        \midrule
        \multirow{6}{*}[0.5em]{5} & \multirow{6}{*}[0.5em]{\includegraphics[width=0.9cm]{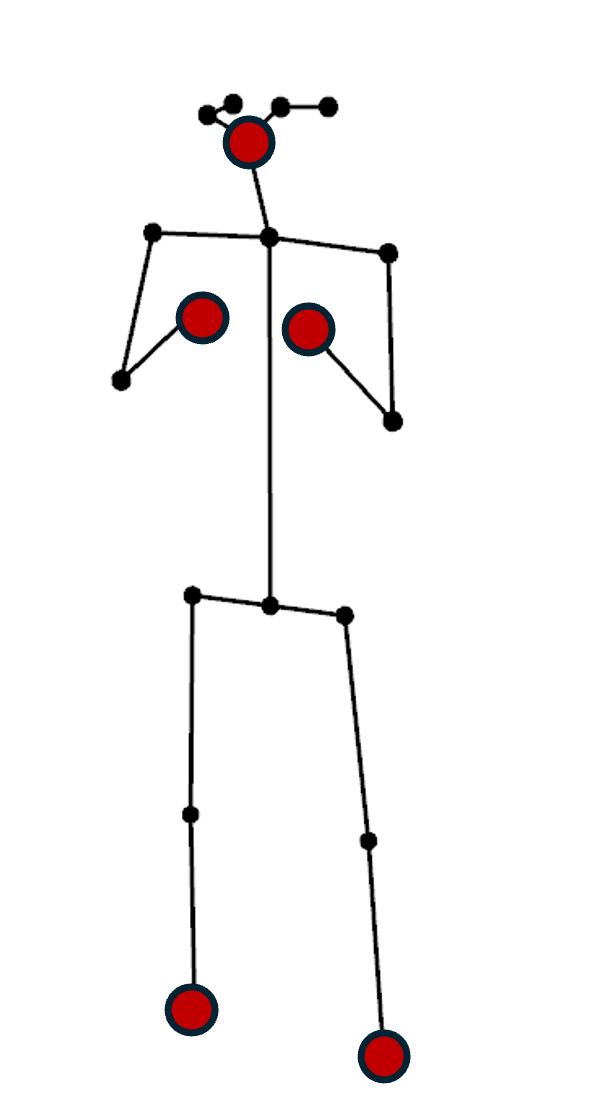}} & \multirow{6}{*}[0.25em]{\includegraphics[width=2.8cm]{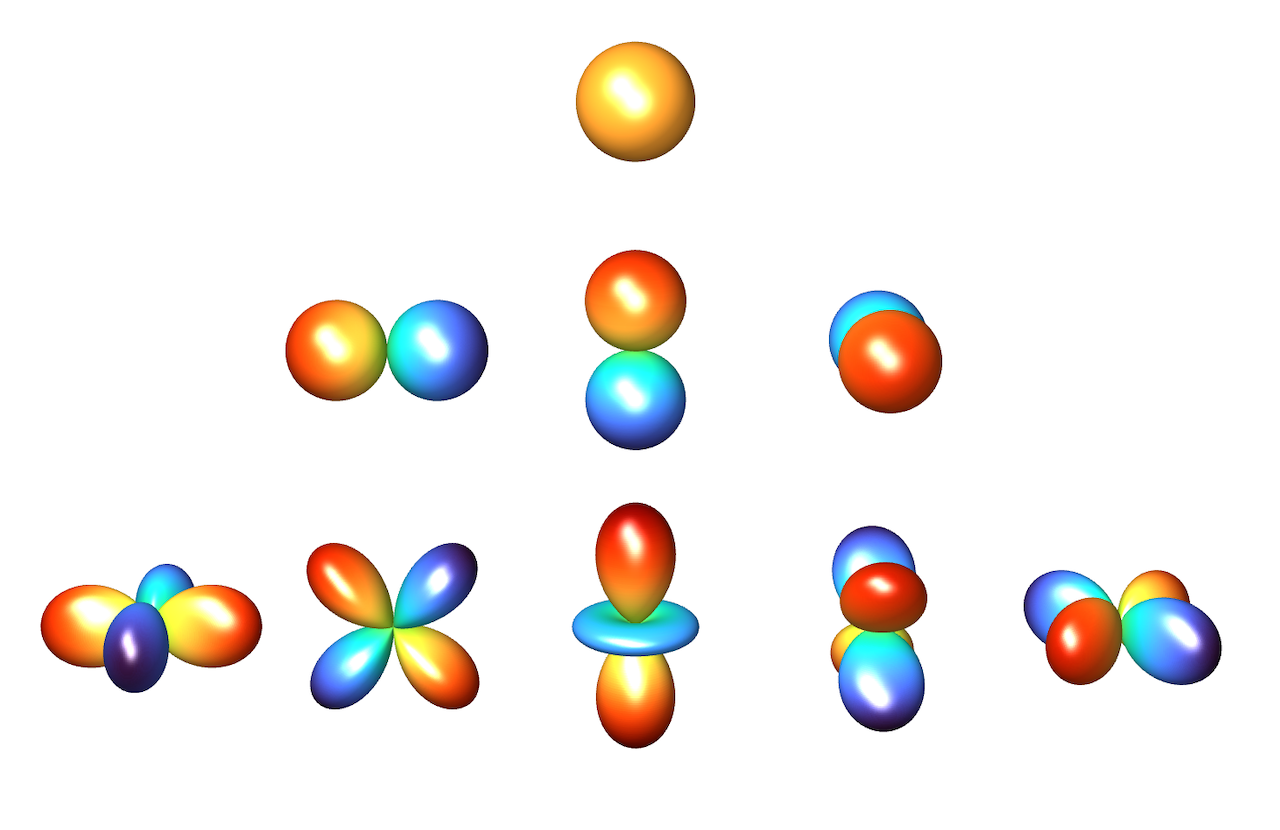}} & \multirow{2}{*}[0.3em]{$0$th} & &\multirow{2}{*}[0.3em]{2.009} & \multirow{2}{*}[0.3em]{1.013}& \multirow{2}{*}[0.3em]{0.334}& &\multirow{2}{*}[0.3em]{4.775} & \multirow{2}{*}[0.3em]{1.225}& \multirow{2}{*}[0.3em]{0.559}\\
         & & & & & & & & & & & \\
         &  &  & \multirow{2}{*}[0.6em]{$1$st} & &\multirow{2}{*}[0.6em]{3.534} & \multirow{2}{*}[0.6em]{0.896}& \multirow{2}{*}[0.6em]{0.322}& &\multirow{2}{*}[0.6em]{7.261} & \multirow{2}{*}[0.6em]{0983}& \multirow{2}{*}[0.6em]{0.480}\\
        & & & & & & & & & & & \\
         &  &  & \multirow{2}{*}[0.9em]{$2$nd} & &\multirow{2}{*}[0.9em]{3.552} & \multirow{2}{*}[0.9em]{0.911}& \multirow{2}{*}[0.9em]{0.323}& &\multirow{2}{*}[0.9em]{7.981} & \multirow{2}{*}[0.9em]{0.977}& \multirow{2}{*}[0.9em]{0.442}\\
         \midrule
         \multirow{6}{*}[0.5em]{9} & \multirow{6}{*}[0.5em]{\includegraphics[width=0.9cm]{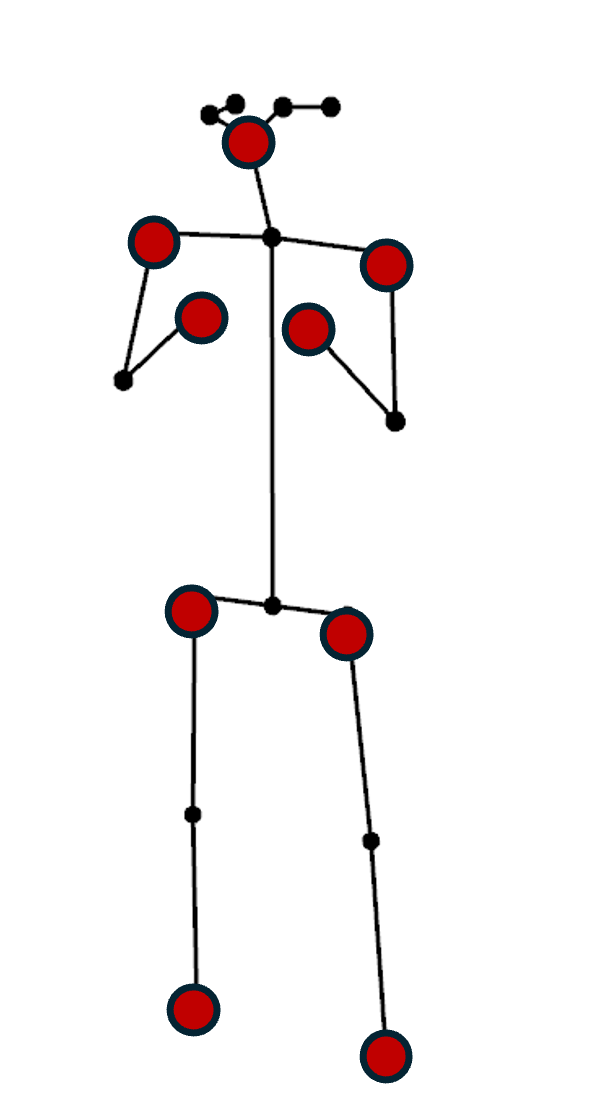}} & \multirow{6}{*}[0.25em]{\includegraphics[width=2.8cm]{figures/SH_2nd.png}} & \multirow{2}{*}[0.3em]{$0$th} & &\multirow{2}{*}[0.3em]{3.059} & \multirow{2}{*}[0.3em]{0.907}& \multirow{2}{*}[0.3em]{0.327}& &\multirow{2}{*}[0.3em]{6.619} & \multirow{2}{*}[0.3em]{1.068}& \multirow{2}{*}[0.3em]{0.496}\\
         & & & & & & & & & & & \\
         &  &  & \multirow{2}{*}[0.6em]{$1$st} & &\multirow{2}{*}[0.6em]{3.600} & \multirow{2}{*}[0.6em]{0.895}& \multirow{2}{*}[0.6em]{0.323}& &\multirow{2}{*}[0.6em]{7.479} & \multirow{2}{*}[0.6em]{0.983}& \multirow{2}{*}[0.6em]{0.467}\\
        & & & & & & & & & & & \\
         &  &  & \multirow{2}{*}[0.9em]{$2$nd} & &\multirow{2}{*}[0.9em]{3.569} & \multirow{2}{*}[0.9em]{0.915}& \multirow{2}{*}[0.9em]{0.323}& &\multirow{2}{*}[0.9em]{8.200} & \multirow{2}{*}[0.9em]{0.952}& \multirow{2}{*}[0.9em]{0.434}\\
         \midrule
         \multirow{6}{*}[0.5em]{12} & \multirow{6}{*}[0.5em]{\includegraphics[width=0.9cm]{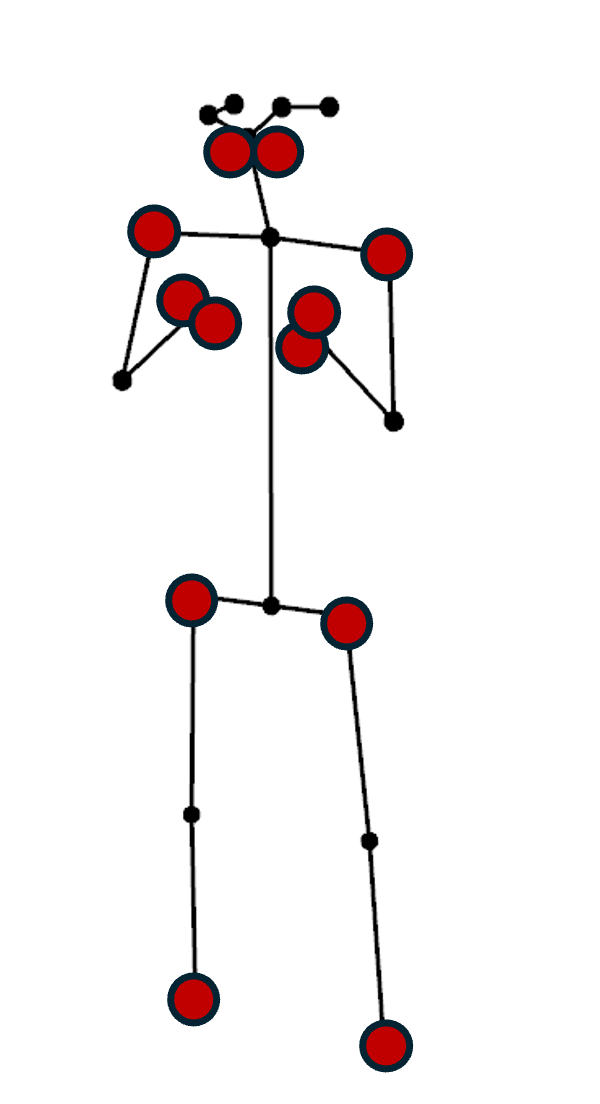}} & \multirow{6}{*}[0.25em]{\includegraphics[width=2.8cm]{figures/SH_2nd.png}} & \multirow{2}{*}[0.3em]{$0$th} & &\multirow{2}{*}[0.3em]{3.020} & \multirow{2}{*}[0.3em]{0.895}& \multirow{2}{*}[0.3em]{0.327}& &\multirow{2}{*}[0.3em]{6.893} & \multirow{2}{*}[0.3em]{1.031}& \multirow{2}{*}[0.3em]{0.472}\\
         & & & & & & & & & & & \\
         &  &  & \multirow{2}{*}[0.6em]{$1$st} & &\multirow{2}{*}[0.6em]{3.616} & \multirow{2}{*}[0.6em]{0.879}& \multirow{2}{*}[0.6em]{0.325}& &\multirow{2}{*}[0.6em]{7.616} & \multirow{2}{*}[0.6em]{0.969}& \multirow{2}{*}[0.6em]{0.466}\\
        & & & & & & & & & & & \\
         &  &  & \multirow{2}{*}[0.9em]{$2$nd} & &\multirow{2}{*}[0.9em]{$3.597$} & \multirow{2}{*}[0.9em]{$0.883$}& \multirow{2}{*}[0.9em]{$0.323$}& &\multirow{2}{*}[0.9em]{$8.448$} & \multirow{2}{*}[0.9em]{$0.943$}& \multirow{2}{*}[0.9em]{$0.417$}\\
         \midrule
         \midrule
        12 & & no primitive offset & $2$nd & & {$3.528$} & {$0.919$} & {$0.321$} &  & {$7.730$} & {$0.998$} & {$0.456$}\\
         \bottomrule
    \end{tabularx}
    \label{tab:num_prim_order}
\end{table}

\subsection{Ablation Study}
\label{subsec:ablation}
In this section we evaluate the impact of the number of acoustic primitives and their harmonic order. Zero-order harmonics are able to model only omidirectional fields, while higher orders allow for increasingly complex radiation patterns. Note that we limit the maximum order to 2 given that PyTorch implementations of spherical wave functions are available only up to the second order. Intuitively, a higher number of acoustic primitives allows for modeling of more complex overall sound fields. We test three configurations of acoustic primitives: 5 primitives: head, L/R hand, L/R foot; 9 primitives: head, L/R hand, L/R foot, L/R shoulder, L/R hip; and 12 primitives where head and hands have two primitives associated with the same key-point (given location offsets these primitives are not bound to be in the same location allowing approximation of a higher order radiation pattern). Results are shown in ~\cref{tab:num_prim_order}. In general, both higher number of primitives and higher primitive order improve the performance as expected. This is especially true for speech. For body sounds on the other hand, increasing from $1$st to $2$nd order does not seem to be beneficial. We also evaluate the model without the primitive offset adjustment. From ~\cref{tab:num_prim_order} we can observe that removing the offset has similar impact as decreasing the number of primitives from 12 to 9, which intuitively makes sense given that repeated primitives collapse to the same key-point location. 

For more experiments, such as ablation on the loss terms and visualizations with different harmonic orders, please refer to the supplementary materials.

\subsection{Qualitative Results}
\begin{figure}[!t]
    \centering
    \includegraphics[width=0.99\textwidth]{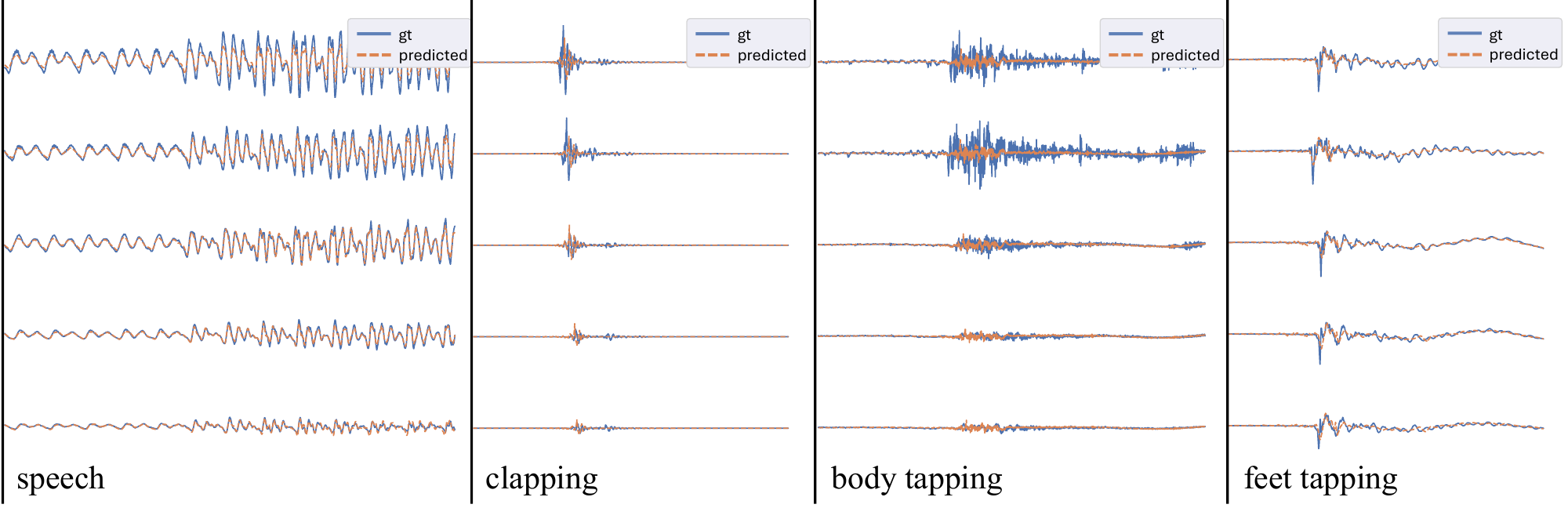}
    \caption{Predicted and ground truth microphone signals at 5 different locations around the dome. We can observe good temporal alignment and good amplitude match except for the low-energy body tapping sound. We recommend zooming in for better visibility.}
    \label{fig:examples_waveform}
\end{figure}

Some sound field visualization examples are shown in \cref{fig:examples_soundfield}. We can observe that the network is able to correctly associate different kinds of sounds to the appropriate acoustic primitives. We can also observe the speech radiation pattern matching the head orientation.
Furthermore, \cref{fig:examples_waveform} shows predicted and ground truth waveforms at different microphone locations. We can observe good temporal alignment and amplitude match for most cases. One exception is the body tapping sound in which the amplitude does not match across different microphones. This may be due to the primitive struggling to match a highly variable radiation pattern.

\section{Conclusion}

We propose a neural rendering system for sound that allows to generate and render 3D sound fields from sparse user input like body pose and headset audio.
We demonstrate the we maintain similar quality to state-of-the-art sound rendering, while improving significantly on speed and soundfield completeness:
our approach is an order of magnitude faster than the approach from~\cite{xudong2023sounding} and is capable of rendering sound in the near-field, \ie close to the transmitter's body, where the previous approach from~\cite{xudong2023sounding} failed.

Moreover, we want to highlight the design similarities to successful neural renderers from computer graphics:
By leveraging an acoustic rendering equation and acoustic primitives, similar to leveraging volumetric primitives in graphical neural rendering, we design a 3D spatial audio system with a conceptual duality to its visual counterpart.
We hope this work will impact sound rendering in 3D settings like computer games and AR/VR.

\noindent\textbf{Limitations.}
Albeit the promising results in terms of quality and efficiency, our approach is still far from broad availability:
model training relies on data collected with a multi-microphone capture stage that is not broadly available.
Future directions need to aim at enabling learning such acoustic scenes with simpler setups, ideally with commodity hardware like smartphones.
Generalization beyond human bodies is another natural extension that emerges from the availability of broader data sources for spatial sound.

\noindent\textbf{Potential Society Impact.}
Ethical and societal risks in this work are low since no data is manipulated in a generative fashion - pure spatialization has little potential for harmful actors.

\bibliographystyle{splncs04}
\bibliography{egbib}

\title{Supplementary Materials for ``Modeling and Driving Human Body Soundfields through Acoustic Primitives''}
\titlerunning{Acoustic Primitives}

\author{}
\institute{}
\authorrunning{C. Huang et al.}

\maketitle

\renewcommand{\thesection}{\Alph{section}}
\setcounter{figure}{5}
\setcounter{equation}{11}
\setcounter{table}{2}

\section{Rendering Audio with Learned Soundfield}
\begin{figure}[h]
    \centering
    \includegraphics[width=0.5\textwidth]{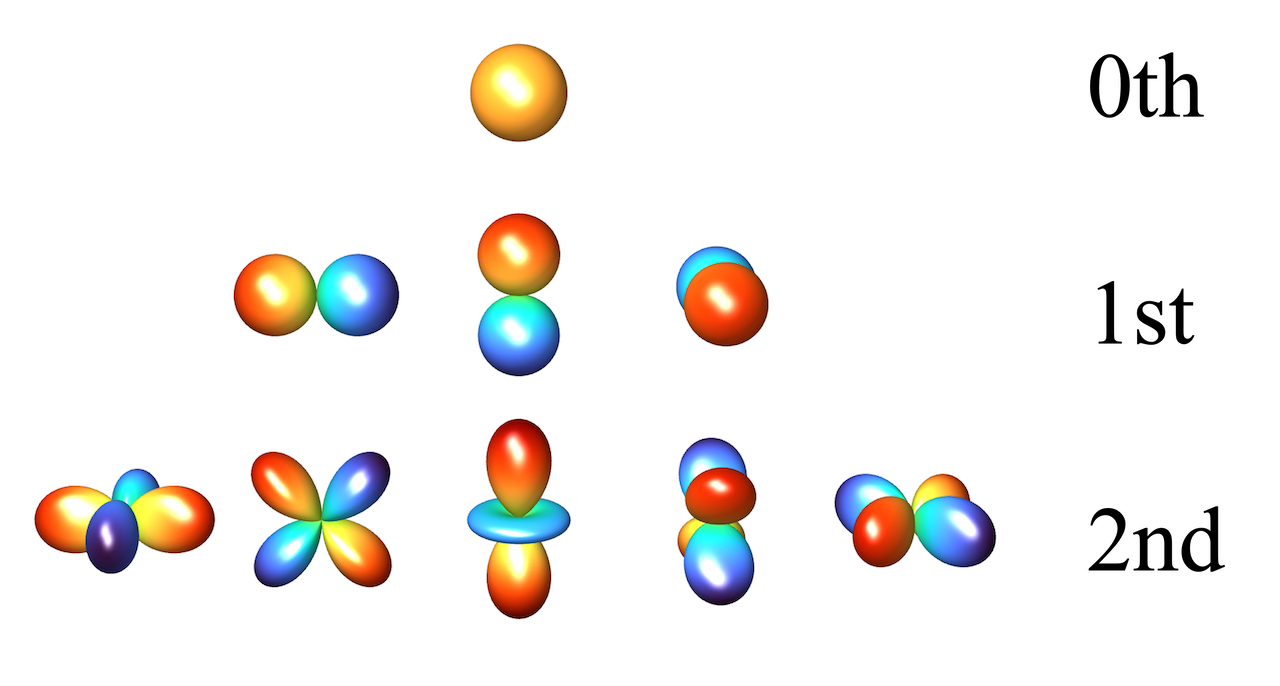}
    \caption{Spherical harmonics up to second order.}
    \label{fig:SH}
\end{figure}

As illustrated in Sec 3.2, the sound pressure, \ie the audio signal, produced by a learned soundfield of order $N$ is given by
\begin{equation}
\begin{split}
\mathbf{w}(r,\theta,\varphi) & = \sum_{n=0}^{N} \sum_{m=-n}^{n} \left({c}_{nm} \cdot h_n (kr_{ref})\right) \cdot \frac{h_n(kr)}{h_n(kr_{ref}) }  \cdot Y_{nm} (\theta, \varphi) \\
& = \sum_{n=0}^{N} \sum_{m=-n}^{n} \Tilde{c}_{nm} \cdot \frac{h_n (kr)}{h_n(kr_{ref}) }  \cdot Y_{nm} (\theta, \varphi),
\end{split}
\label{eq:primitives_}
\end{equation}
here, $k = 2\pi f / v_{sound}$ is the corresponding wavenumber; $Y_{nm} (\theta, \varphi)$ represents the spherical harmonic of order $n$ and degree $m$, which is 
\begin{equation}
    Y_{nm} (\theta, \varphi) \equiv \sqrt{\frac{2n+1}{4\pi}\frac{(n-m)!}{(n+m)!}}P_{nm}(\cos\theta)e^{im\varphi},
\end{equation}
$P_{nm}(z)$ is an associated Legendre polynomial, and $h_n (kr)$ is $n$th-order spherical Hankel functions. All the functions are implemented with PyTorch and, therefore are fully differentiable. In this paper, we choose spherical harmonics up to second order and showcase them in \cref{fig:SH}.

\begin{figure}[t]
    \centering
    \includegraphics[width=0.99\textwidth]{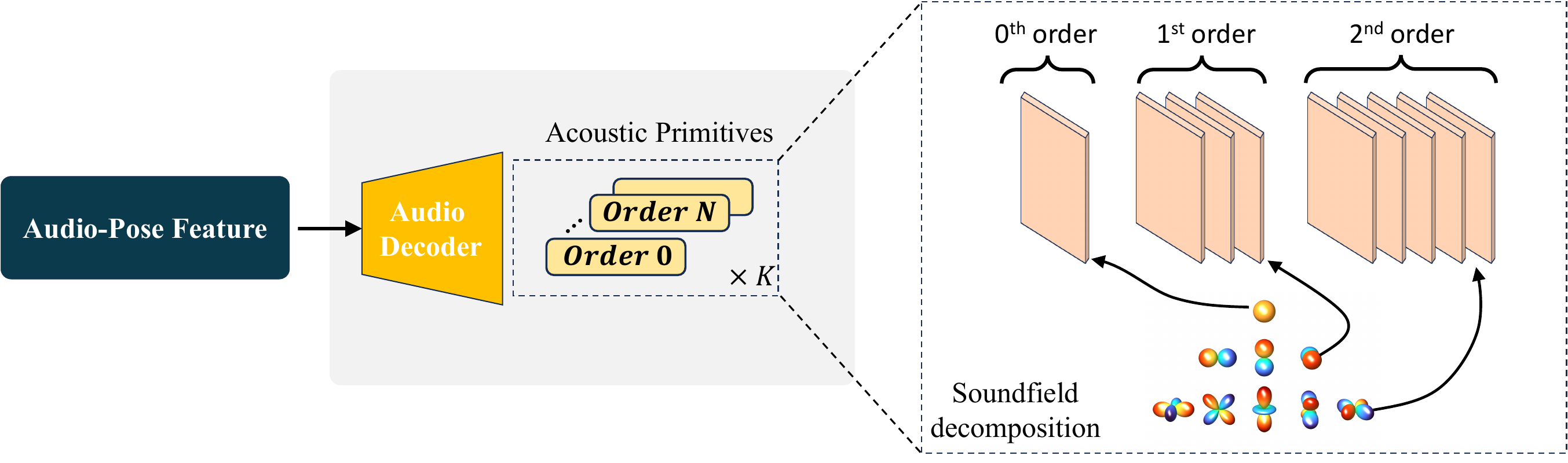}
    \caption{Decomposition of learned soundfield representation and its connection to second order spherical harmonics.}
    \label{fig:decomposition}
\end{figure}

Similarly to $Y_{nm} (\theta, \varphi)$, the learned soundfield, \ie acoustic primitive, can also be decomposed into a series of spherical harmonics functions, each representing a different spatial component of the soundfield. We demonstrate the decomposition process in \cref{fig:decomposition}. Our learned soundfield representation is enforced to express the same spatial information as spherical harmonics because each predicted acoustic primitive has $(N+1)^2$ channels, which is equivalent to the number of spherical harmonics.

\begin{figure}[h]
    \centering
    \includegraphics[width=0.9\textwidth]{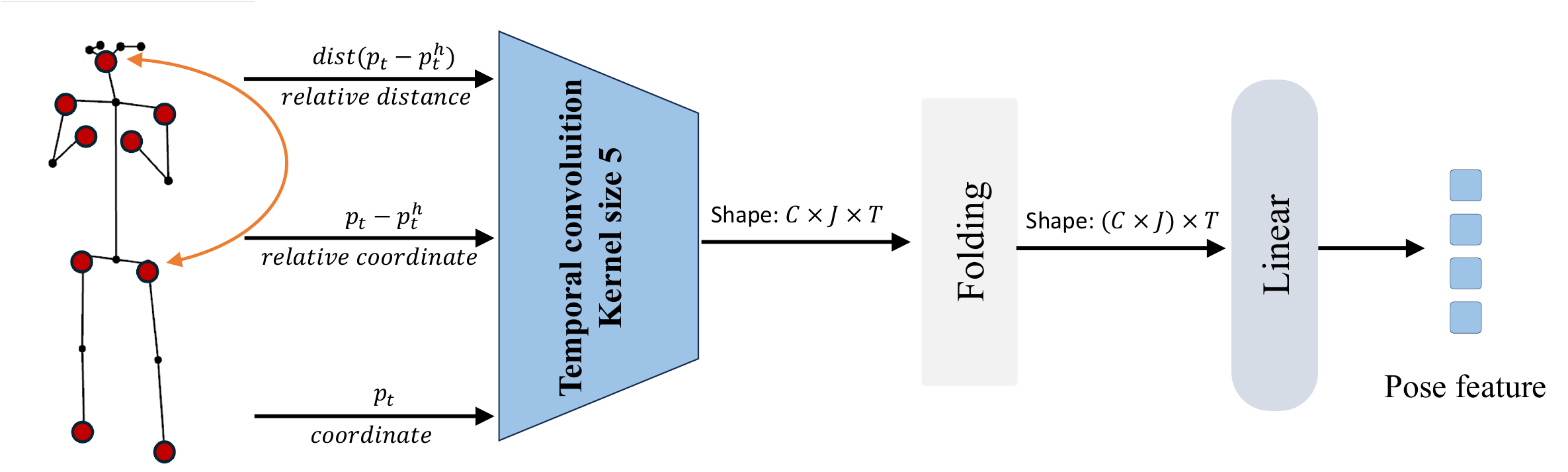}
    \caption{Illsutration on pose feature encoding, we consider the absolute coordinates and the relative coordinates/distances (to the headset) of body joints.}
    \label{fig:pose_encoding}
    \vspace{-5mm}
\end{figure}

\section{Pose Feature Encoding}
To capture these rich spatial cues, we employ a pose encoder that processes the input pose sequence.
This sequence, denoted as  $\mathbf{p}_{1:T_p}$,  contains the 3D coordinates of body joints for each frame $\mathbf{p}_{t} \in \mathbb{R} ^ {J \times 3}$. However, since these coordinates are captured from a third-person perspective, they might not fully capture the spatial relationship relevant to the audio, where the sound originates from the body but is received at the headset. To address this, we enhance the pose input by selecting the head joint $\mathbf{p}^{h}_{t}$ as an anchor and calculating relative coordinates and Euclidean distances. This extended pose input $[\mathbf{p}_{t}, \mathbf{p}_{t} - \mathbf{p}^{h}_{t}, dist(\mathbf{p}_{t}-\mathbf{p}^h_{t})] \in \mathbb{R} ^ {J \times 7}$, consisting of original coordinates, relative coordinates to the head, and distance from the head, provides the pose encoder with a more comprehensive understanding of the body's spatial relationship with the sound. Details are shown in \cref{fig:pose_encoding}.

\section{More Ablations}

\begin{figure}[H]
    \centering
    \vspace{-5mm}
    \includegraphics[width=0.9\textwidth]{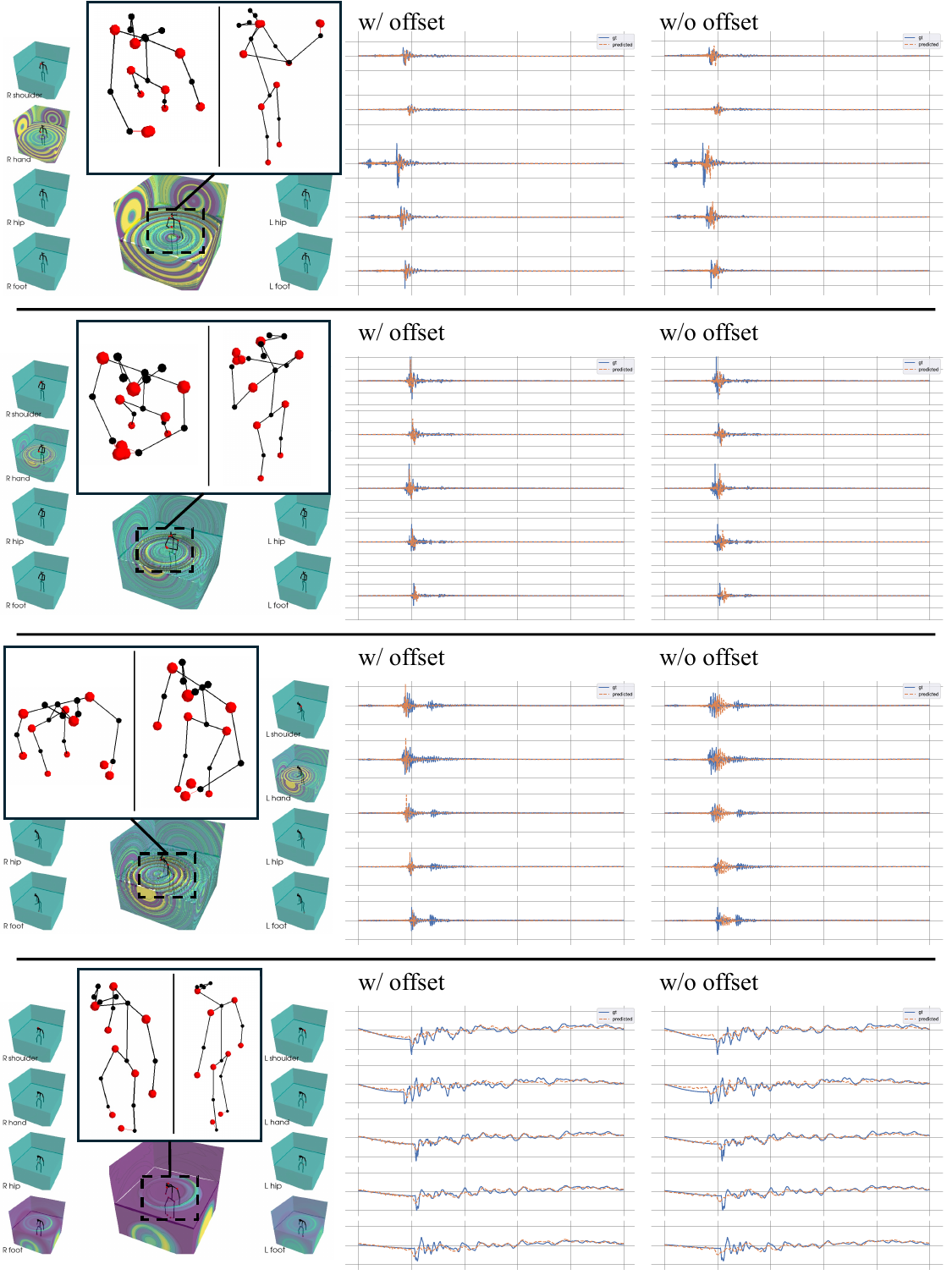}
    \caption{Ablation on learned acoustic primitive offsets. 
    }
    \label{fig:offset_examples}
    \vspace{-5mm}
\end{figure}

\noindent\textbf{Visualization of primitive offsets.} In \cref{fig:offset_examples}, we observe a \textbf{time delay} between the predicted and GT audios for the \textbf{model without offsets}, likely caused by inaccurate primitive coordinates.  In contrast, our model with learned offsets mitigates this issue, resulting in a closer match to the ground truth audio. Also, we visualize the sound fields generated by our framework for different primitives after applying the learned offsets.
We observe that the learned offsets generally match location where we would expect the source of sound to be given the particular sound event such as snap, clap, or footstep. 

\begin{figure}[H]
    \centering
    \vspace{-5mm}
    \includegraphics[width=0.8\textwidth]{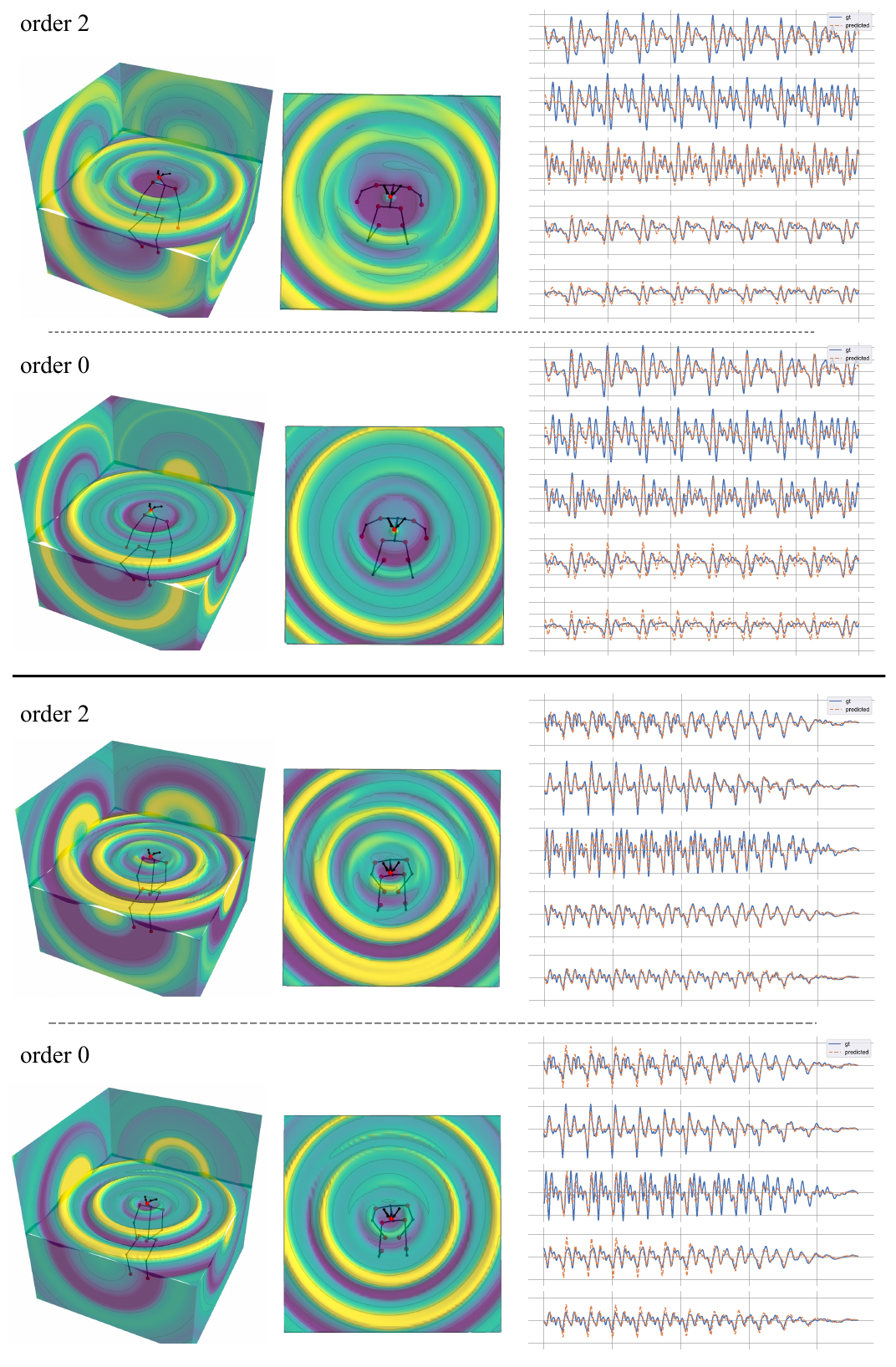}
    \caption{Ablation on different harmonic orders.}
    \label{fig:order_examples}
    \vspace{-5mm}
\end{figure}

\noindent\textbf{Visualization of different harmonic order.} \cref{fig:order_examples} illustrates the impact of sound field order on the accuracy of predicted audio. As shown, the model's prediction using a $2$nd-order sound field exhibits \textbf{a closer match to the GT audio in terms of amplitude}. This is because higher-order harmonics offer finer spatial rendering capabilities, allowing the model to capture more precise directional details of the sound. In contrast, the predicted $0$th-order sound field is omnidirectional, meaning it radiates sound equally in all directions. This limitation hinders its ability to encode specific spatial information, resulting in less accurate audio amplitude prediction.

\noindent\textbf{Ablation on the choices of loss function.} 
In ~\cref{tab:loss}, we conduct an ablation study to investigate the effectiveness of each loss term in the total loss function (Eq. (11) in the main paper). We remove each loss term from the total loss $\mathcal{L}_{total}$ one at a time. The results show that  including $\mathcal{L}_{cts}$ improves the overall performance on both speech and non-speech data, and combining all the loss terms yields the best or second-best performance across different metrics for both speech and non-speech data and generally the best performance on average over the metrics.
\begin{table}[!t]
    \caption{Ablation study on loss function. Each loss term is removed from the total loss function one at a time. The best and second-best results are highlighted in \textcolor{green}{green} and \textcolor{blue}{blue}, respectively.}
    \centering
    \footnotesize
    \begin{tabularx}{1\textwidth}{lXrrrXrrr}
        \toprule
                        & & \multicolumn{3}{c}{non-speech}                                        & & \multicolumn{3}{c}{speech} \\
                            \cmidrule(lr){3-5}                                                        \cmidrule(lr){7-9}
        Loss   & & SDR $\uparrow$    & amplitude $\downarrow$    & phase $\downarrow$    & & SDR $\uparrow$    & amplitude $\downarrow$    & phase $\downarrow$ \\
        \midrule
        $\mathcal{L}_{total}$  &  &\textcolor{green}{3.597} & \textcolor{blue}{0.883} & \textcolor{blue}{0.323} & & \textcolor{blue}{8.448} & 0.943 & \textcolor{blue}{0.417} \\ 
        w/o $\mathcal{L}_{amp}$ & & 2.099 & 1.179 & \textcolor{green}{0.303} & & 8.370 & 1.038 & \textcolor{green}{0.381} \\ 
        w/o $\mathcal{L}_{ri}$ & &\textcolor{blue}{3.579} & 0.894 & 0.325 & & \textcolor{green}{8.492} & \textcolor{blue}{0.933} & 0.418 \\ 
        w/o $\mathcal{L}_{s\ell1}$ & & 3.338 & \textcolor{green}{0.821} & 0.330 & & 7.722 & \textcolor{green}{0.915} & 0.491 \\ 
        w/o $\mathcal{L}_{cts}$ & & 3.523 & 0.903 & 0.324 & & 8.239 & 0.938 & 0.422 \\ \bottomrule
    \end{tabularx}
    \label{tab:loss}
\end{table}

\section{Demo Video} 
We have prepared a supplementary video to visually demonstrate the capabilities of our method in spatial audio rendering. The video showcases a full-body avatar producing correctly spatialized binaural audio corresponding to various actions and interactions using our trained model. In particular, the input of the audio system is a single channel mono audio that contains the mixture of all sounds being made. Our model can render them with the correct spatial locations using wearer's body pose. This means that the wearer can \textit{clap left, clap right, applaud, snap around,} and the sounds will be appropriately positioned. Additionally, the system works with \textit{objects that the wearer may be using}, such as an egg shaker.

\section*{Acknowledgement}
The authors would like to thank Frank Yu for engineering work on the VR demo.

\end{document}